\documentclass{sig-alternate-05-2015}
\pdfoutput=1 

\def \mypapertitle {Learning to Match Using Local and Distributed Representations of Text for Web Search}

\makeatletter
\def\@copyrightspace{\relax}
\makeatother

\usepackage{booktabs}
\usepackage[inline]{enumitem}

\usepackage[config=altsf,caption=false]{subfig}
\usepackage{graphicx}
\usepackage{fixltx2e}

\usepackage{url,times}
\usepackage{microtype}

\usepackage{amsmath}
\usepackage{amssymb}
\usepackage{multirow}
\usepackage[square,comma,numbers,sort&compress,sectionbib]{natbib}
\usepackage{color}
\usepackage[usenames,dvipsnames]{xcolor}
\usepackage[normalem]{ulem}
\usepackage{authblk}
\usepackage{fdiaz}
\usepackage{tikz}

\newcommand\mybox[2][]{\tikz[overlay]\node[fill=blue!20,inner sep=2pt, anchor=text, rectangle, rounded corners=1mm,#1] {#2};\phantom{#2}}

\newif\ifanon
\anonfalse

\newcommand{\ndrm}[0]{f}
\newcommand{\ndrmLocal}[0]{\ndrm_{\ell}}
\newcommand{\ndrmDistributed}[0]{\ndrm_{d}}

\newcommand{\strlen}[0]{n}

\newcommand{\qterm}[0]{\fdvec{q}}
\newcommand{\query}[0]{\fdmat{Q}}
\newcommand{\qlen}[0]{\strlen_{\qterm}}

\newcommand{\dterm}[0]{\fdvec{d}}
\newcommand{\doc}[0]{\fdvec{D}}
\newcommand{\dlen}[0]{\strlen_{\dterm}}

\newcommand{\interactionLocal}[0]{\fdmat{X}}
\newcommand{\interactionDistributed}[0]{\tilde{\interactionLocal}}

\newcommand{\inputWordDimension}[0]{m}
\newcommand{\inputWordDimensionLocal}[0]{\inputWordDimension_{\ell}}
\newcommand{\inputWordDimensionDistributed}[0]{\inputWordDimension_{d}}

\newcommand{\numConvolutionsLocal}[0]{c}

\newcommand{\convolutionOutputLocal}[0]{\fdmat{Z}}
\newcommand{\convolutionWeightsLocal}[0]{\fdvec{W}}

\newcommand{\reldoc}[0]{\doc^*}
\newcommand{\negset}[0]{\fdset{N}}

\newcommand{\numneg}[0]{\text{N}}
\newcommand{\maxgraph}[0]{\text{G}}
\newcommand{\cdssmmaxwords}[0]{\text{T}}

\newcommand{\distributedQuery}[0]{\tilde{\query}}
\newcommand{\distributedDoc}[0]{\tilde{\doc}}

\begin{document}

\setcopyright{acmcopyright}

\doi{10.475/123_4}

\isbn{123-4567-24-567/08/06}

\conferenceinfo{Under review for WWW '17}{April 3--7, 2017, Perth, Australia}
\acmPrice{\$15.00}

\title{\mypapertitle}

\numberofauthors{3}

\ifanon
\alignauthor
\else
\renewcommand\Authfont{\fontsize{12}{14.4}}
\renewcommand\Affilfont{\fontsize{9}{10.8}}
\author[1,2]{Bhaskar Mitra\thanks{The author is a part-time PhD student at UCL.} }
\author[1]{Fernando Diaz}
\author[1]{Nick Craswell}
\affil[1]{Microsoft, \{bmitra, fdiaz, nickcr\}@microsoft.com}
\affil[2]{University College London, \{bhaskar.mitra.15\}@ucl.ac.uk}
\fi

\maketitle
\begin{abstract}

Models such as latent semantic analysis and those based on neural embeddings learn \emph{distributed} representations of text, and match the query against the document in the latent semantic space. In traditional information retrieval models, on the other hand, terms have discrete or \emph{local} representations, and the relevance of a document is determined by the exact matches of query terms in the body text. We hypothesize that matching with distributed representations complements matching with traditional local representations, and that a combination of the two is favorable. We propose a novel document ranking model composed of two separate deep neural networks, one that matches the query and the document using a local representation, and another that matches the query and the document using learned distributed representations. The two networks are jointly trained as part of a single neural network. We show that this combination or `duet' performs significantly better than either neural network individually on a Web page ranking task, and also significantly outperforms traditional baselines and other recently proposed models based on neural networks.

\end{abstract}

\section{Introduction}
\label{sec:intro}

\begin{figure}[t]
\centering
\subfigure[Local model]{\noindent
{\small
    \parbox{0.90\columnwidth}{%
\textcolor[gray]{0.750000}{The}
\mybox[fill=green!18]{\textcolor[gray]{0.250000}{President}}
\textcolor[gray]{0.750000}{of}
\textcolor[gray]{0.750000}{the}
\mybox[fill=green!40]{\textcolor[gray]{0.250000}{United}}
\mybox[fill=green!22]{\textcolor[gray]{0.250000}{States}}
\textcolor[gray]{0.750000}{of}
\textcolor[gray]{0.750000}{America}
\textcolor[gray]{0.750000}{(POTUS)}
\textcolor[gray]{0.750000}{is}
\textcolor[gray]{0.750000}{the}
\textcolor[gray]{0.750000}{elected}
\textcolor[gray]{0.750000}{head}
\textcolor[gray]{0.750000}{of}
\textcolor[gray]{0.750000}{state}
\textcolor[gray]{0.750000}{and}
\textcolor[gray]{0.750000}{head}
\textcolor[gray]{0.750000}{of}
\textcolor[gray]{0.750000}{government}
\textcolor[gray]{0.750000}{of}
\textcolor[gray]{0.750000}{the}
\mybox[fill=green!5]{\textcolor[gray]{0.250000}{United}}
\mybox[fill=green!10]{\textcolor[gray]{0.250000}{States.}}
\textcolor[gray]{0.750000}{The}
\mybox[fill=green!5]{\textcolor[gray]{0.250000}{president}}
\textcolor[gray]{0.750000}{leads}
\textcolor[gray]{0.750000}{the}
\textcolor[gray]{0.750000}{executive}
\textcolor[gray]{0.750000}{branch}
\textcolor[gray]{0.750000}{of}
\textcolor[gray]{0.750000}{the}
\textcolor[gray]{0.750000}{federal}
\textcolor[gray]{0.750000}{government}
\textcolor[gray]{0.750000}{and}
\textcolor[gray]{0.750000}{is}
\textcolor[gray]{0.750000}{the}
\textcolor[gray]{0.750000}{commander}
\textcolor[gray]{0.750000}{in}
\textcolor[gray]{0.750000}{chief}
\textcolor[gray]{0.750000}{of}
\textcolor[gray]{0.750000}{the}
\textcolor[gray]{0.750000}{United}
\mybox[fill=green!2]{\textcolor[gray]{0.250000}{States}}
\textcolor[gray]{0.750000}{Armed}
\textcolor[gray]{0.750000}{Forces.}
\textcolor[gray]{0.750000}{Barack}
\textcolor[gray]{0.750000}{Hussein}
\textcolor[gray]{0.750000}{Obama}
\textcolor[gray]{0.750000}{II}
\textcolor[gray]{0.750000}{(born}
\textcolor[gray]{0.750000}{August}
\textcolor[gray]{0.750000}{4,}
\textcolor[gray]{0.750000}{1961)}
\textcolor[gray]{0.750000}{is}
\textcolor[gray]{0.750000}{an}
\textcolor[gray]{0.750000}{American}
\textcolor[gray]{0.750000}{politician}
\textcolor[gray]{0.750000}{who}
\textcolor[gray]{0.750000}{is}
\textcolor[gray]{0.750000}{the}
\textcolor[gray]{0.750000}{44th}
\textcolor[gray]{0.750000}{and}
\textcolor[gray]{0.750000}{current}
\mybox[fill=green!5]{\textcolor[gray]{0.250000}{President}}
\textcolor[gray]{0.750000}{of}
\textcolor[gray]{0.750000}{the}
\textcolor[gray]{0.750000}{United}
\textcolor[gray]{0.750000}{States.}
\textcolor[gray]{0.750000}{He}
\textcolor[gray]{0.750000}{is}
\textcolor[gray]{0.750000}{the}
\textcolor[gray]{0.750000}{first}
\textcolor[gray]{0.750000}{African}
\textcolor[gray]{0.750000}{American}
\textcolor[gray]{0.750000}{to}
\textcolor[gray]{0.750000}{hold}
\textcolor[gray]{0.750000}{the}
\textcolor[gray]{0.750000}{office}
\textcolor[gray]{0.750000}{and}
\textcolor[gray]{0.750000}{the}
\textcolor[gray]{0.750000}{first}
\mybox[fill=green!1]{\textcolor[gray]{0.250000}{president}}
\textcolor[gray]{0.750000}{born}
\textcolor[gray]{0.750000}{outside}
\textcolor[gray]{0.750000}{the}
\textcolor[gray]{0.750000}{continental}
\mybox[fill=green!3]{\textcolor[gray]{0.250000}{United}}
\mybox[fill=green!1]{\textcolor[gray]{0.250000}{States.}}
}}
}

\subfigure[Distributed model]{\noindent
{\small
    \parbox{0.90\columnwidth}{%
\textcolor[gray]{0.750000}{The}
\textcolor[gray]{0.750000}{President}
\textcolor[gray]{0.750000}{of}
\textcolor[gray]{0.750000}{the}
\mybox[fill=green!2]{\textcolor[gray]{0.250000}{United}}
\textcolor[gray]{0.750000}{States}
\textcolor[gray]{0.750000}{of}
\textcolor[gray]{0.750000}{America}
\textcolor[gray]{0.750000}{(POTUS)}
\textcolor[gray]{0.750000}{is}
\textcolor[gray]{0.750000}{the}
\textcolor[gray]{0.750000}{elected}
\textcolor[gray]{0.750000}{head}
\mybox[fill=green!8]{\textcolor[gray]{0.250000}{of}}
\textcolor[gray]{0.750000}{state}
\textcolor[gray]{0.750000}{and}
\textcolor[gray]{0.750000}{head}
\textcolor[gray]{0.750000}{of}
\textcolor[gray]{0.750000}{government}
\textcolor[gray]{0.750000}{of}
\mybox[fill=green!1]{\textcolor[gray]{0.250000}{the}}
\mybox[fill=green!9]{\textcolor[gray]{0.250000}{United}}
\mybox[fill=green!2]{\textcolor[gray]{0.250000}{States.}}
\mybox[fill=green!2]{\textcolor[gray]{0.250000}{The}}
\mybox[fill=green!3]{\textcolor[gray]{0.250000}{president}}
\textcolor[gray]{0.750000}{leads}
\textcolor[gray]{0.750000}{the}
\mybox[fill=green!3]{\textcolor[gray]{0.250000}{executive}}
\mybox[fill=green!3]{\textcolor[gray]{0.250000}{branch}}
\mybox[fill=green!26]{\textcolor[gray]{0.250000}{of}}
\mybox[fill=green!19]{\textcolor[gray]{0.250000}{the}}
\mybox[fill=green!10]{\textcolor[gray]{0.250000}{federal}}
\textcolor[gray]{0.750000}{government}
\mybox[fill=green!24]{\textcolor[gray]{0.250000}{and}}
\mybox[fill=green!6]{\textcolor[gray]{0.250000}{is}}
\mybox[fill=green!5]{\textcolor[gray]{0.250000}{the}}
\mybox[fill=green!10]{\textcolor[gray]{0.250000}{commander}}
\mybox[fill=green!21]{\textcolor[gray]{0.250000}{in}}
\mybox[fill=green!2]{\textcolor[gray]{0.250000}{chief}}
\textcolor[gray]{0.750000}{of}
\mybox[fill=green!1]{\textcolor[gray]{0.250000}{the}}
\mybox[fill=green!8]{\textcolor[gray]{0.250000}{United}}
\textcolor[gray]{0.750000}{States}
\mybox[fill=green!11]{\textcolor[gray]{0.250000}{Armed}}
\textcolor[gray]{0.750000}{Forces.}
\mybox[fill=green!4]{\textcolor[gray]{0.250000}{Barack}}
\mybox[fill=green!29]{\textcolor[gray]{0.250000}{Hussein}}
\mybox[fill=green!36]{\textcolor[gray]{0.250000}{Obama}}
\mybox[fill=green!40]{\textcolor[gray]{0.250000}{II}}
\mybox[fill=green!5]{\textcolor[gray]{0.250000}{(born}}
\textcolor[gray]{0.750000}{August}
\textcolor[gray]{0.750000}{4,}
\textcolor[gray]{0.750000}{1961)}
\mybox[fill=green!5]{\textcolor[gray]{0.250000}{is}}
\mybox[fill=green!19]{\textcolor[gray]{0.250000}{an}}
\textcolor[gray]{0.750000}{American}
\textcolor[gray]{0.750000}{politician}
\mybox[fill=green!11]{\textcolor[gray]{0.250000}{who}}
\mybox[fill=green!12]{\textcolor[gray]{0.250000}{is}}
\textcolor[gray]{0.750000}{the}
\textcolor[gray]{0.750000}{44th}
\textcolor[gray]{0.750000}{and}
\textcolor[gray]{0.750000}{current}
\textcolor[gray]{0.750000}{President}
\textcolor[gray]{0.750000}{of}
\mybox[fill=green!7]{\textcolor[gray]{0.250000}{the}}
\mybox[fill=green!15]{\textcolor[gray]{0.250000}{United}}
\textcolor[gray]{0.750000}{States.}
\mybox[fill=green!4]{\textcolor[gray]{0.250000}{He}}
\textcolor[gray]{0.750000}{is}
\textcolor[gray]{0.750000}{the}
\textcolor[gray]{0.750000}{first}
\textcolor[gray]{0.750000}{African}
\textcolor[gray]{0.750000}{American}
\mybox[fill=green!0]{\textcolor[gray]{0.250000}{to}}
\mybox[fill=green!10]{\textcolor[gray]{0.250000}{hold}}
\mybox[fill=green!23]{\textcolor[gray]{0.250000}{the}}
\textcolor[gray]{0.750000}{office}
\textcolor[gray]{0.750000}{and}
\textcolor[gray]{0.750000}{the}
\textcolor[gray]{0.750000}{first}
\textcolor[gray]{0.750000}{president}
\mybox[fill=green!2]{\textcolor[gray]{0.250000}{born}}
\mybox[fill=green!12]{\textcolor[gray]{0.250000}{outside}}
\mybox[fill=green!23]{\textcolor[gray]{0.250000}{the}}
\mybox[fill=green!23]{\textcolor[gray]{0.250000}{continental}}
\mybox[fill=green!6]{\textcolor[gray]{0.250000}{United}}
\mybox[fill=green!1]{\textcolor[gray]{0.250000}{States.}}
}}
}
\caption{Visualizing the drop in each model's retrieval score by individually removing each of the passage terms for the query ``united states president''. Darker green signifies bigger drop. The local model uses only exact term matches.  The distributed model uses matches based on a learned representation.}
\label{fig:wordimportance}
\end{figure}

Neural text embedding models have recently gained significant popularity for both natural language processing (NLP) and information retrieval (IR) tasks. In IR, a significant number of these works have focused on word embeddings \cite{mitra2016desm, nalisnick2016improving, ganguly2015word, diaz2016query, grbovic2015context, grbovic2015search, roy2016using, zheng2015learning} and modelling short-text similarities \cite{huang2013learning, shen2014learning, shen2014latent, palangi2016deep, hu2014convolutional, severyn2015learning}. In traditional Web search, the query consists of only few terms but the body text of the documents typically has at least few hundred sentences. In the absence of click information, such as for newly-published or infrequently-visited documents, the body text can be a useful signal to determine the relevance of the document for the query. Therefore, extending existing neural text representation learning approaches to long body text for document ranking is an important challenge in IR. However, as was noted during a recent workshop \cite{craswellneu}, in spite of the recent surge in interests towards applying deep neural network (DNN) models for retrieval, their success on ad-hoc retrieval tasks has been rather limited. Some recent papers \cite{pang2016study, roy2016using} report worse  performance of neural embedding models when compared to traditional term-based approaches, such as BM25 \cite{robertson2009probabilistic}.

Traditional IR approaches consider terms as discrete entities. The relevance of the document to the query is estimated based on, amongst other factors, the number of matches of query terms in the document, the parts of the document in which the matches occur, and the proximity between the matches. In contrast, latent semantic analysis (LSA) \cite{deerwester1990indexing}, probabilistic latent semantic analysis (PLSA) \cite{hofmann1999probabilistic} and latent Dirichlet allocation (LDA) \cite{blei2003latent, wei2006lda} learn low-dimensional vector representations of terms, and match the query against the document in the latent semantic space. Retrieval models can therefore be classified based on what representations of text they employ at the point of matching the query against the document. At the point of match, if each term is represented by a unique identifiers (\emph{local} representation \cite{hinton1984distributed}) then the query-document relevance is a function of the pattern of occurrences of the exact query terms in the document. However, if the query and the document text is first projected into a continuous latent space, then it is their distributed representations that are compared. Along these lines, \citet{guodeep} classify recent DNN models for short-text matching as either \emph{interaction}-focused \cite{lu2013deep, hu2014convolutional, pang2016text, guodeep} or \emph{representation}-focused \cite{huang2013learning, shen2014learning, shen2014latent, hu2014convolutional, severyn2015learning}. They claim that IR tasks are different from NLP tasks, and that it is more important to focus on exact matching for the former and on learning text embeddings for the latter. \citet{mitra2016desm}, on the other hand, claim that models that compare the query and the document in the latent semantic space capture a different sense of relevance than models that focus on exact term matches, and therefore the combination of the two is more favorable. Our work is motivated by the latter intuition that it is important to match the query and the document using both local and distributed representations of text. We propose a novel ranking model comprised of two separate DNNs that model query-document relevance using local and distributed representations, respectively. The two DNNs, referred to henceforth as the \emph{local model} and the \emph{distributed model}, are jointly trained as part of a single neural network, that we name as a \emph{duet} architecture because the two networks co-operate to achieve a common goal. Figure \ref{fig:wordimportance} demonstrates how each subnetwork models the same document given a fixed query.  While the local model captures properties like exact match position and proximity, the distributed model detects property synonyms (e.g. `Obama'), related terms (e.g. `federal'), and even wellformedness of content (e.g. `the', `of')\footnote{While surprising, this last property is important for detecting quality web content \cite{yun:web-clarity}.}.

In this paper, we show that the duet of the two DNNs not only outperforms the individual local and distributed models, but also demonstrate large improvements over traditional baselines and other recently proposed models based on DNNs on the document ranking task. Unlike many other work, our model significantly outperforms classic IR approaches by using a DNN to learn text representation.

DNNs are known for requiring significant training data, and most of the state-of-the-art performances achieved by these deep models are in areas where large scale corpora are available for training \cite{lecun2015deep, jozefowicz2016exploring}. Some of the lack of positive results from neural models in the area of ad-hoc retrieval is likely due to the scarce public availability of large quantity of training data necessary to learn effective representations of text. In Section \ref{sec:discussion}, we will present some analysis on the effect of training data on the performance of these DNN models. In particular, we found that--unsurprisingly--the performance of the distributed model improves drastically in the presence of more data. Unlike some previous work \cite{huang2013learning, shen2014learning, shen2014latent} that train on clickthrough data with randomly sampled documents as negative examples, we train our model on human-judged labels. Our candidate set for every query consists of documents that were retrieved by the commercial search engine Bing, and then labelled by crowdsourced judges. We found that training with the documents that were rated irrelevant by the human judges as the negative examples is more effective than randomly sampling them from the document corpus. To summarize, the key contributions of this work are:

\begin{enumerate}
    \itemsep-2pt
    \item We propose a novel duet architecture for a model that jointly learns two deep neural networks focused on matching using local and distributed representations of text, respectively.
     \item We demonstrate that this architecture out-performs state-of-the-art neural and traditional non-neural baselines.
    \item We demonstrate that training with documents judged as irrelevant as the negative examples is more effective than randomly sampling them from the corpus.
\end{enumerate}

\section{Desiderata of Document Ranking}
\label{sec:motivation}
Before describing our ranking model, we first present three properties found across most effective retrieval systems.  We will then operationalize these in our architecture in Section \ref{sec:model}.

First, \emph{exact term matches} between the query and the document are fundamental to all information retrieval models \cite{fang:axoimatic}.  Traditional IR models such as BM25 \cite{robertson2009probabilistic} are based on counts of exact matches of query terms in the document text. They can be employed with minimal (or no) need for training data, sometimes directly on new tasks or corpora. Exact matching can be particularly important when the query terms are new or rare. For example, if new documents appear on the Web with the television model number `SC32MN17' then BM25 can immediately retrieve these pages containing precisely that model number without adjusting any parameters of the ranking model. A good ranking model needs to take advantage of exact matches to perform reliably on fresh and rare queries.

\begin{figure}
\centering
\includegraphics[width=.5\textwidth]{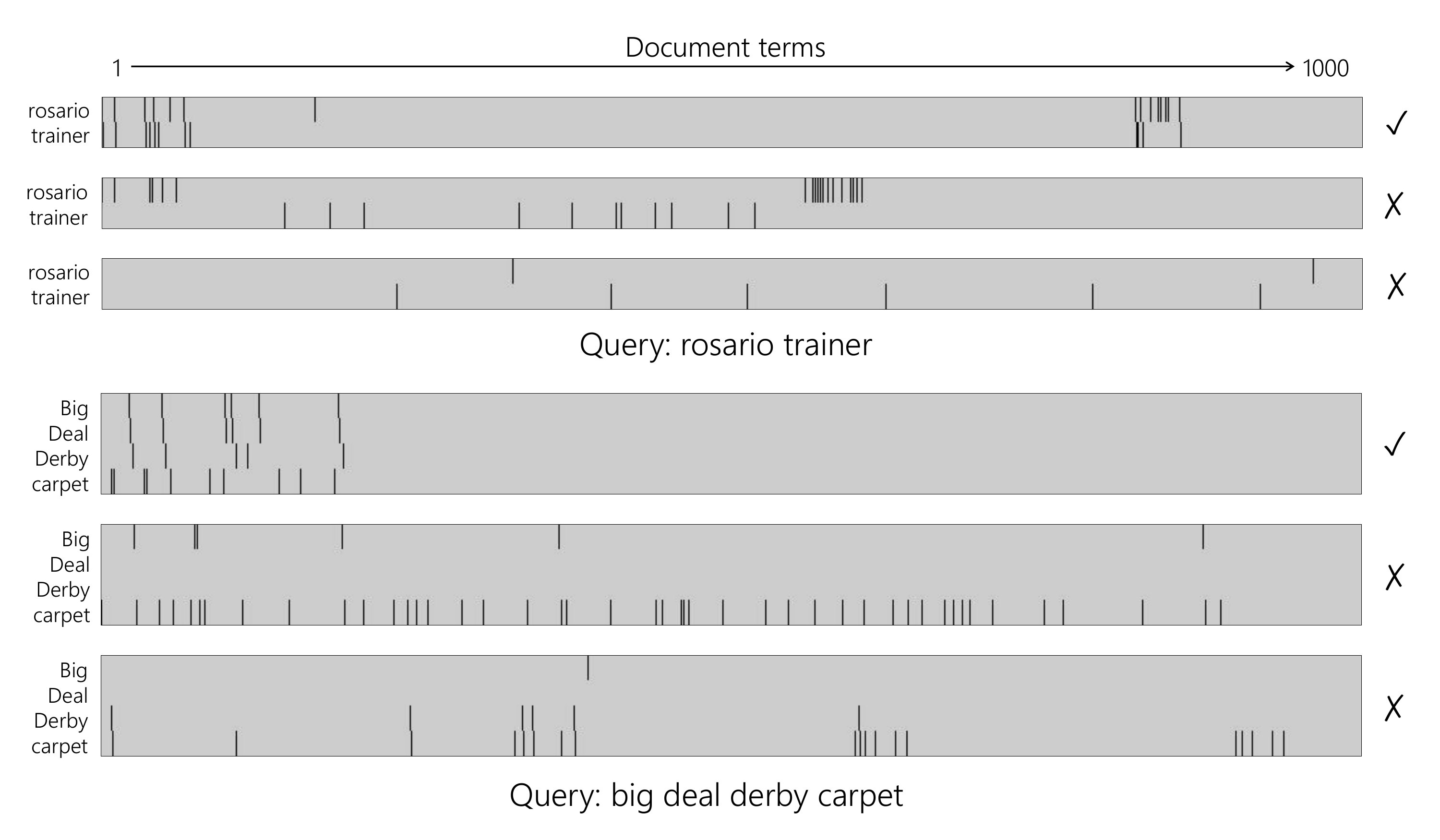}
\caption{Visualizing patterns of query term matches in documents. Query terms are laid out along the vertical axis, and the document terms along the horizontal axis. The short vertical lines correspond to exact matches between pairs of query and document terms. For both of the above queries, the first document was rated relevant by a human judge and the following two as irrelevant.}
\label{fig:interactionmatrix}
\end{figure}

Second, \emph{match positions} of query terms in the document not only reflect where potentially relevant parts of the document are localized (e.g. title, first paragraph, closing paragraph) but also how clustered individual query term matches are with each other.  Figure \ref{fig:interactionmatrix} shows the position of matches on two different queries and a sample of relevant and non-relevant documents.  In the first query, we see that query term matches in the relevant document are much more clustered than in the non-relevant documents.  We observe this behavior in the second query but also notice that the clustered matches are localized near the beginning of the relevant document.  Match proximity serves as a foundation for effective methods such as sequential dependence models \cite{metzler2005markov}.

Finally,  \emph{inexact term matches} between the query and the document refer to techniques for addressing the vocabulary mismatch problem.  The main disadvantage of term matching is that related terms are ignored, so when ranking for the query `Australia' then only the term frequency of `Australia' is considered, even though counting terms like `Sydney' and `koala' can be good positive evidence. \citet{robertson2004understanding} pointed out that under the probabilistic model of IR there is, in fact, no good justification for ignoring the non-matching terms in the document. Furthermore, \citet{mitra2016desm} demonstrated that a distributed representation based retrieval model that considered \emph{all} document terms is able to better distinguish between a passage that is truly relevant to the query ``Cambridge'' from a passage on a different topic (e.g., giraffes) with artificially injected occurrences of the term ``Cambridge''. Any IR model that considers the distribution of non-matching terms is therefore likely to benefit from this additional evidence, and be able to tell ``Cambridge'' apart from ``an African even-toed ungulate mammal''.

In practice, the most effective techniques leverage combinations of these techniques.  Dependence models combine exact matching with proximity \cite{metzler2005markov}.  LDA-based document models  combine exact matching with inexact matching \cite{wei2006lda}.  Query hypergraphs capture all three \cite{bendersky:hypergraphs}.  Our method also combines these techniques but, unlike  prior work, jointly learns all free parameters of the different components.

\section{The Duet Architecture}
\label{sec:model}

\begin{figure}[!ht]
\centering
\includegraphics[width=0.90\columnwidth]{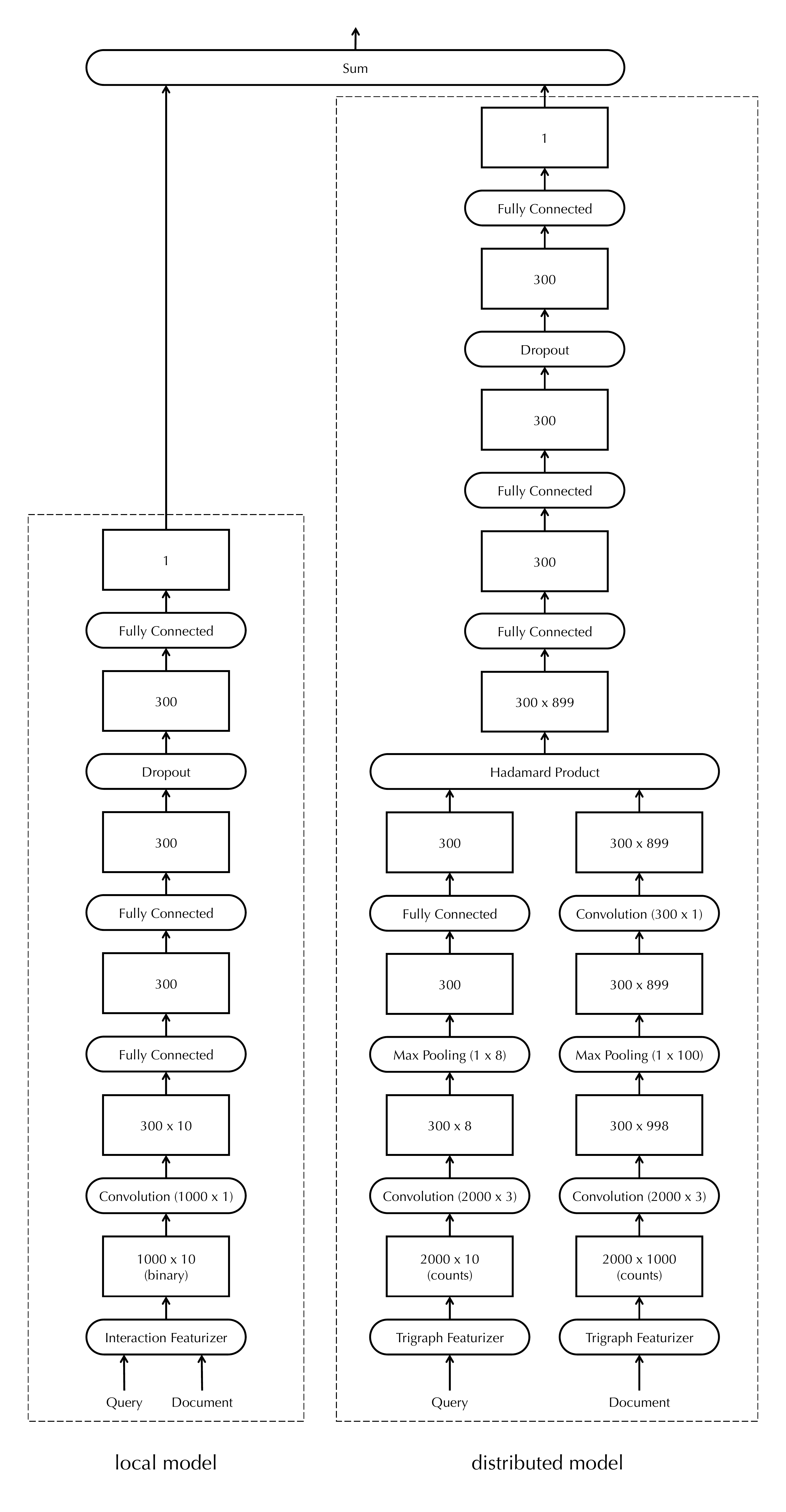}
\caption{The duet architecture composed of the local model (left) and the distributed model (right). The local sub-network takes an interaction matrix of query and document terms as input, whereas the distributed sub-network learns embeddings of the query and document text before matching.}
\label{fig:architecture}
\end{figure}

Figure \ref{fig:architecture} provides a detailed schematic view of the duet architecture. The distributed model projects the query and the document text into an embedding space before matching, while the local model operates over an interaction matrix comparing every query term to every document term. The final score under the duet setup is the sum of scores from the local and the distributed networks,
\begin{align}
\ndrm(\query,\doc) &= \ndrmLocal(\query,\doc) + \ndrmDistributed(\query,\doc)\label{eqn:ndrm_disccont}
\end{align}
where  both the query and the document are considered as ordered list of terms, $\query = [\qterm_1,\ldots,\qterm_{\qlen}]$ and  $\doc = [\dterm_1,\ldots,\dterm_{\dlen}]$.  Each query term $\qterm$ and document term $\dterm$ is a $\inputWordDimension\times 1$ vector where $\inputWordDimension$ is the input representation of the text (e.g. the number of terms in the vocabulary for the local model).  We fix the length of the inputs across all queries and documents such that we consider only the first 10 terms in the query and the first 1000 in the document.  If the either is shorter than the target dimension, the input vectors are padded with zeros. The truncation of the document body text to the first 1000 terms is performed only for our model variants. For all the neural and non-neural baseline models we consider the full body text.

\subsection{Local Model}
\label{sec:modellocal}

The local model estimates document relevance based on patterns of exact matches of query terms in the document.  To this end, each term is represented by its one-hot encoding in a $\inputWordDimensionLocal$-dimensional space, where $\inputWordDimensionLocal$ is the size of the vocabulary.  The model then generates the $\dlen\times \qlen$ binary matrix $\interactionLocal=\fdtrans{\doc}\query$, capturing every exact match (and position) of query terms in the document. This interaction matrix, without the zero-padding, is analogous to the visual representation of term matches in Figure \ref{fig:interactionmatrix}, and therefore captures both exact term matches and match positions. It is also similar to the indicator matching matrix proposed previously by \citet{pang2016text}. While interaction matrix $\interactionLocal$ perfectly captures every query term match in the document, it does not retain information about the actual terms themselves. Therefore, the local model cannot learn term-specific properties from the training corpus, nor model interactions between dissimilar terms.

The interaction matrix $\interactionLocal$ is first passed through a convolutional layer with $\numConvolutionsLocal$ filters, a kernel size of $\dlen \times 1$, and a stride of $1$. The output $\convolutionOutputLocal_{i}$ corresponding to the $i^{th}$ convolutional window over $\interactionLocal$ is a function of the match between the $\qterm_i$ term against the whole document,
\begin{align}
\label{eqn:ndrm_localconvo}
\convolutionOutputLocal_{i} = \tanh\left(\fdtrans{\interactionLocal_{i}}\convolutionWeightsLocal\right)
\end{align}
where $\interactionLocal_i$ is row $i$ of $\interactionLocal$, $\tanh$ is performed elementwise,  and the $\dlen\times\numConvolutionsLocal $ matrix $\convolutionWeightsLocal$ contains the learnable parameters of the convolutional layer. The output $\convolutionOutputLocal$ of the convolutional layer is a matrix of dimension $\numConvolutionsLocal \times \qlen$. We use a filter size ($\numConvolutionsLocal$) of 300 for all the evaluations reported in this paper. The output of the convolutional layer is then passed through two fully-connected layers, a dropout layer, and a final fully-connected layer that produces a single real-valued output. All the nodes in the local model uses the hyperbolic tangent function for non-linearity.

\subsection{Distributed Model}
\label{sec:modeldistrib}

The distributed model learns dense lower-dimensional vector representations of the query and the document text, and then computes the positional similarity between them in the learnt embedding space. Instead of a one-hot encoding of terms, as in the local model, we use a character $n$-graph based representation of each term in the query and document. Our $n$-graph based input encoding is motivated by the trigraph encoding proposed by \citet{huang2013learning}.  For each term, we count all of the $n$-graphs present for $1\leq n\leq \maxgraph$.  We then use this $n$-graph frequency vector of length $\inputWordDimensionDistributed$ to represent the term.

Instead of directly computing the interaction between the $\inputWordDimensionDistributed \times \qlen$ matrix $\query$ and  the $\inputWordDimensionDistributed \times \dlen$ matrix $\doc$, we first learn a series of  nonlinear transformations to the character-based input. For both query and document, the first step is convolution. The $\inputWordDimensionDistributed \times 3$ convolution window  has filter size of 300. It projects 3 consecutive terms to a 300 dimensional vector, then takes a stride by 1 position, and projects the next 3 terms, and so on. For the query, the convolution step generates a tensor of dimensions $300 \times 8$. For document it generates one of dimensions $300 \times 998$.

Following this, we conduct a max-pooling step. For the query the pooling kernel dimensions are $1 \times 8$. For document it is $1 \times 100$. As a result, we get one $300 \times 1$ matrix $\distributedQuery$ for the query and a $300 \times 899$ matrix $\distributedDoc$ for the document. The document matrix $\distributedDoc$ can be interpreted as $899$ separate embeddings, each corresponding to different equal-sized spans of text within the document. Our choice of a window-based max-pooling strategy, instead of global max-pooling as employed by CDSSM \cite{shen2014learning}, is motivated by the fact that the window-based approach allows the model to distinguish between matches in different parts of the document. As posited in Section \ref{sec:motivation}, a model that is aware of match positions may be more suitable when dealing with long documents, especially those containing mixture of many topics.

The output of the max-pooling layer for the query is then passed through a fully-connected layer. For the document, the $300 \times 899$ dimensional matrix output is operated on by another convolutional layer with filter size of 300, kernel dimensions of $300 \times 1$, and a stride of 1. The combination of these convolutional and max-pooling layers enable the distributed model to learn suitable representations of text for effective inexact matching.

In order to perform the matching, we conduct the element-wise or Hadamard product between the embedded document matrix and the extended or broadcasted query embedding,

\begin{align}
\interactionDistributed&= (\underbrace{\distributedQuery\ldots\distributedQuery}_{\text{899 times}})\circ\distributedDoc
\end{align}

After this, we pass the matrix through fully connected layers, and a dropout layer until we arrive at a single score. Similar to the local model, we use hyperbolic tangent function here for non-linearity.
\subsection{Optimization}
\label{sec:training}
Each training sample consists of a query $\query$, a relevant document $\reldoc$ and a set of irrelevant documents $\negset=\{\doc_0,\ldots,\doc_{\numneg}\}$. We use a softmax function to compute the posterior probability of the positive document given a query based on the  score.

\begin{align}
p(\reldoc|\query) &= \frac{\exp(\ndrm(\query,\reldoc))}{\sum_{\doc \in \negset}\exp(\ndrm(\query,\doc))}\label{eqn:ndrm_softmax}
\end{align}

and we maximize the log likelihood $\log p(\reldoc|\query)$ using  stochastic gradient descent.

\section{Materials and Methods}
\label{sec:experiments}

We conducted three experiments to test: 
\begin{enumerate*}[label=(\arabic*)]
    \item the effectiveness of our duet model compared to the local and distributed models separately, and
    \item the effectiveness of our duet model compared to existing baselines for content-based web ranking,
    \item the effectiveness of training with judged negative documents compared to random negative documents.
\end{enumerate*}
In this section, we detail our experiment setup and baseline implementations. 

\subsection{Data}
\label{sec:data}

The training dataset consisted of 199,753 instances in the format described in Section \ref{sec:experiment:training}. The queries in the training dataset were randomly sampled from Bing's search logs from a period between January, 2012 and September, 2014. Human judges rated the documents on a five-point scale (\emph{perfect}, \emph{excellent}, \emph{good}, \emph{fair} and \emph{bad}).  The document body text was retrieved from Bing's Web document index. We used proprietary parsers for extracting the body text from the raw HTML content. All the query and the document text were normalized by down-casing and removing all non-alphanumeric characters.

We considered two different test sets, both sampled from Bing search logs. The \emph{weighted} set consisted of queries sampled according their frequency in the search logs.  As a result, frequent queries were well-represented in this dataset.  Queries were sampled between October, 2014 and December, 2014.   The \emph{unweighted} set consisted of queries sampled uniformly from the entire population of unique queries.  The queries in this samples removed the bias toward  popular queries found in the weighted set.  The unweighted queries were sampled between January, 2015 and June, 2015. 

Because all of our datasets were derived from sampling real query logs and because queries will naturally repeat, there was some overlap in queries between the training and testing sets.  Specifically, 14\% of the testing queries in the weighted set occurred in the training set, whereas only 0.04\% of the testing queries in the unweighted set occurred in the training set.  We present both results for those who may be in environments with repeated queries (as is common in production search engines) and for those who may be more interested in cold start situations or tail queries. Table \ref{tbl:testsets} summarizes statistics for the two test sets. 

\begin{table}[t]
\begin{center}
\caption{Statistics of the three test sets randomly sampled from Bing's search logs. The candidate documents are generated by querying Bing and then rated using human judges.}
\label{tbl:testsets}

\begin{tabular}{lrrr}
  \toprule
 & queries & documents & \( \frac{\textrm{documents}}{\textrm{query}} \) \\ 
 \midrule
 training & 199,753& 998,765 & 5\\
 \\
 weighted & 7,741 & 171,302 & 24.9 \\
 unweighted & 6,808 & 71,722 & 10.6 \\
 \bottomrule
\end{tabular}
\end{center}
\end{table}

\subsection{Training}
\label{sec:experiment:training}

Besides the architecture (Figure \ref{fig:architecture}), our model has the following free parameters: the maximum order of the character-based representation for the distributed model ($\maxgraph$), the number of negative documents to sample at training time ($\numneg$), the dropout rate, and the learning rate. 

We used a maximum order of five for our character $n$-graphs in the distributed model.  Instead of using the full 62,193,780-dimensional vector, we only considered  top 2,000 most popular $n$-graphs, resulting 36 unigraphs (a-z and 0-9), 689 bigraphs, 1149 trigraphs, 118 4-graphs, and eight 5-graphs.

When training our model (Section \ref{sec:training}), we sampled four negative documents for every one relevant document.  More precisely, for each query we generated a maximum of one training sample of each form,
\begin{enumerate*}[label=(\arabic*)]
    \item One \emph{excellent} document with four \emph{fair} documents
    \item One \emph{excellent} document with four \emph{bad} documents
    \item One \emph{good} document with four \emph{bad} documents
\end{enumerate*}.
Pilot experiments showed that treating documents judged as \emph{fair} or \emph{bad} as the negative examples resulted in significantly better performance, than when the model was trained with randomly sampled negatives. For training, we discarded all documents rated as \emph{perfect} because a large portion of them fall under the navigational intent, which can be better satisfied by historical click based ranking signals.

The dropout rate and the learning rate were set to 0.20 and 0.01, respectively, based on a validation set. We implemented our model using CNTK \cite{yu2014introduction} and trained the model with stochastic gradient descent based optimization (with automatic differentiation) on a single GPU. It was necessary to use a small minibatch size of 8 to fit the whole data in GPU memory.\footnote{We will publicly release a CNTK implementation of our model by the time of publication of this paper.}
 
\subsection{Baselines}
\label{sec:baselines}

Our baselines capture the individual properties we outlined in Section \ref{sec:motivation}.  Exact term matching is effectively performed by many classic information retrieval models.  We used the Okapi BM25 \cite{robertson2009probabilistic} and query likelihood (QL) \cite{ponte1998language} models as representative of this class of model.  We used Indri\footnote{\url{http://www.lemurproject.org/indri/}} for indexing and retrieval.

Match positions are handled by substantially fewer models.  Metzler's dependence model (DM) \cite{metzler2005markov} provides an inference network approach to modeling term proximity.  We used the Indri implementation for our experiments.

Inexact term matching received both historic and modern treatments in the literature.  Deerwester \emph{et al.} originally presented latent semantic analysis (LSA) \cite{deerwester1990indexing} as a method for addressing vocabulary mismatch by projecting words and documents into a lower-dimension latent space.  
The dual embedding space model (DESM) \cite{mitra2016desm, nalisnick2016improving} computes a document relevance score by comparing every term in the document with every query term using pre-trained word embeddings. We used the same pre-trained word embeddings dataset that the authors made publicly available online for download\footnote{\url{https://www.microsoft.com/en-us/download/details.aspx?id=52597}}. These embeddings, for approximately 2.8M words, were previously trained on a corpus of Bing queries. In particular, we use the $\text{DESM}_{\text{IN-OUT}}$ model, which was reported to have the best performance on the retrieval task, as a baseline in this paper.
Both the deep structured semantic model (DSSM) \cite{huang2013learning} and its convolutional variant CDSSM \cite{shen2014learning}  consider only the document title for matching with the query. While some papers have reported negative performances for title-based DSSM and CDSSM on the \emph{ad hoc} document retrieval tasks \cite{pang2016study, guodeep}, we included document-based variants appropriately retrained on the same set of positive query and document pairs as our model. As with the original implementation we choose the irrelevant documents for training by randomly sampling from the document corpus. For the CDSSM model, we concatenated the trigraph hash vectors of the first $\cdssmmaxwords$ terms of the body text followed by a vector that is a sum of the trigraph hash vectors for the remaining terms. The choice of $\cdssmmaxwords$ was constrained by memory requirements, and we pick 499 for our experiments.

The DRMM model \cite{guodeep} uses a DNN to perform term matching, with few hundred parameters, over histogram-based features. The histogram features, computed using exact term matching and pre-trained word embeddings based cosine similarities, ignoring the actual position of matches. We implemented the $\text{DRMM}_{\text{LCH} \times \text{IDF}}$ variant of the model on CNTK \cite{yu2014introduction} using word embeddings trained on a corpus of 341,787,174 distinct sentences randomly sampled from Bing's Web index, with a corresponding vocabulary of 5,108,278 words. Every training sample for our model was turned into four corresponding training samples for DRMM, comprised of the query, the positive document, and each one of the negative documents. This guaranteed that both models observed the exact same pairs of positive and negative documents during training. We adopted the same loss function as proposed by Guo \emph{et al.}

\subsection{Evaluation}
\label{sec:experiments:evaluation}
All evaluation and empirical analysis used the normalized discounted cumulative gain (NDCG) metric computed at positions one and ten \cite{NDCG}.  All performance metrics were averaged over queries for each run.  Whenever testing for significant differences in performance, we used the paired $t$-test with a Bonferroni correction.

\section{Results}
\label{sec:results}

\begin{table}[!t]
\begin{center}
\caption{Performance on test data.
All duet runs significantly outperformed our local and distributed model ($p<0.05$).  
All duet runs also outperformed non-neural and neural baselines. The difference between the duet model and the best performing baseline per dataset and position (italics) is statistically significant ($p < 0.05$). The best NDCG performance on each dataset and position is highlighted in bold.}
\label{tbl:results-main}
\subtable[weighted]{
\begin{tabular}{llcc}
  \toprule
 & & NDCG$\mathcal{@}$1 & NDCG$\mathcal{@}$10 \\
 \midrule
\multicolumn{4}{l}{\textbf{Non-neural baselines}}\\
LSA  & & 22.4 & 44.2 \\
BM25  & & 24.2 & 45.5\\
DM  & & 24.7 & 46.2 \\
QL  & & 24.6 & 46.3 \\
 \midrule
\multicolumn{4}{l}{\textbf{Neural baselines}}\\ 
DRMM  & & 24.3 & 45.2 \\
DSSM  & & 25.8 & 48.2 \\
CDSSM  & & \emph{27.3} & 48.2 \\
DESM  & & 25.4 & \emph{48.3} \\
 \midrule
\multicolumn{4}{l}{\textbf{Our models}}\\ 
Local model & & 24.6 & 45.1 \\
Distributed model & & 28.6 & 50.5 \\
Duet model & & \textbf{32.2} & \textbf{53.0}\\
  \bottomrule
\end{tabular}
}

\subtable[unweighted]{
\begin{tabular}{llcc}
  \toprule
 & & NDCG$\mathcal{@}$1 & NDCG$\mathcal{@}$10 \\
 \midrule
\multicolumn{4}{l}{\textbf{Non-neural baselines}}\\
LSA  & & 31.9 & 62.7 \\
BM25  & & 34.9 & 63.3 \\
DM  & & 35.0 & 63.4 \\
QL  & & 34.9 & 63.4 \\
 \midrule
\multicolumn{4}{l}{\textbf{Neural baselines}}\\ 
DRMM  & & \emph{35.6} & \emph{65.1} \\
DSSM  & & 34.3 & 64.4 \\
CDSSM  & & 34.3 & 64.0 \\
DESM  & & 35.0 & 64.7 \\
 \midrule
\multicolumn{4}{l}{\textbf{Our models}}\\ 
Local model & & 35.0 & 64.4 \\
Distributed model & & 35.2 & 64.9 \\
Duet model & & \textbf{37.8} & \textbf{66.4} \\
  \bottomrule
\end{tabular}
}
\end{center}
\end{table}

Table \ref{tbl:results-main} reports NDCG based evaluation results on two test datasets for our model and all the baseline models. Our main observation is that the duet model performs significantly better than the individual local and distributed models. This supports our underlying hypothesis that matching in a latent semantic space can complement exact term matches in a document ranking task, and hence a combination of the two is more appropriate. Note that the NDCG numbers for the local and the distributed models correspond to when these DNNs are trained individually, but for the `duet' the two DNNs are trained together as part of a single neural network. 

Among the baseline models, including both traditional and neural network based models, CDSSM  and DESM  achieve the highest NDCG at position one and ten, respectively, on the weighted test set. On the unweighted test set DRMM  is our best baseline model at both rank positions. The duet model demonstrates significant improvements over all these baseline models on both test sets and at both NDCG positions. 

We also tested our independent local and distributed models against their conceptually closest baselines.  Because our local model captures both matching and proximity, we compared performance to dependence models (DM).  While the performance in terms of NDCG@1 is statistically indistinguishable, both of the NDCG@10 results are statistically significant ($p<0.05$).  We compared our distributed model to the best neural model for each test set and metric.  We found no statistically significant difference except for NDCG@10 for the weighted set.  

\begin{figure}
\centering
\subfigure[Weighted set]{\includegraphics[width=1.5in]{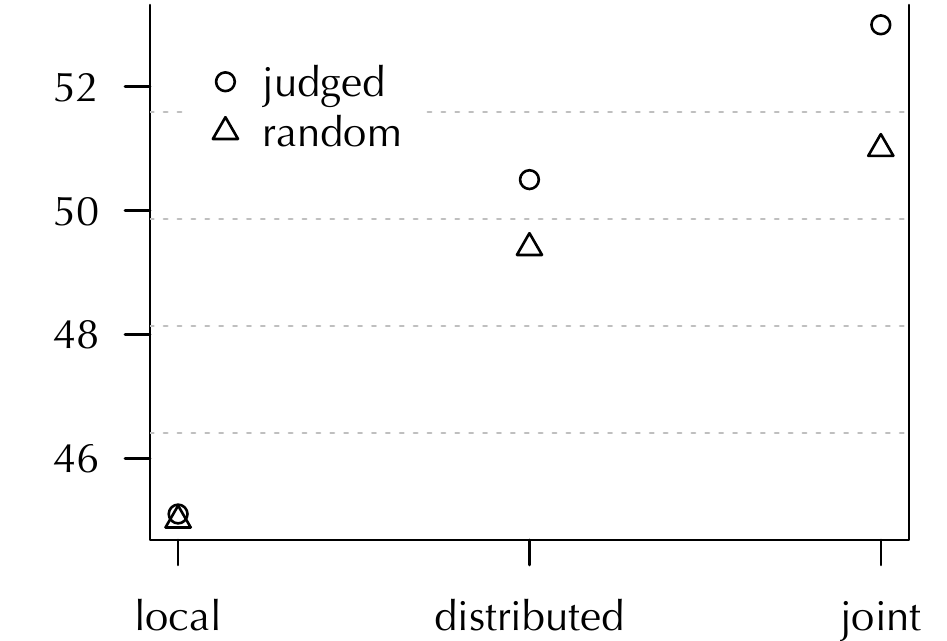}}\hfill\subfigure[Unweighted set]{\includegraphics[width=1.5in]{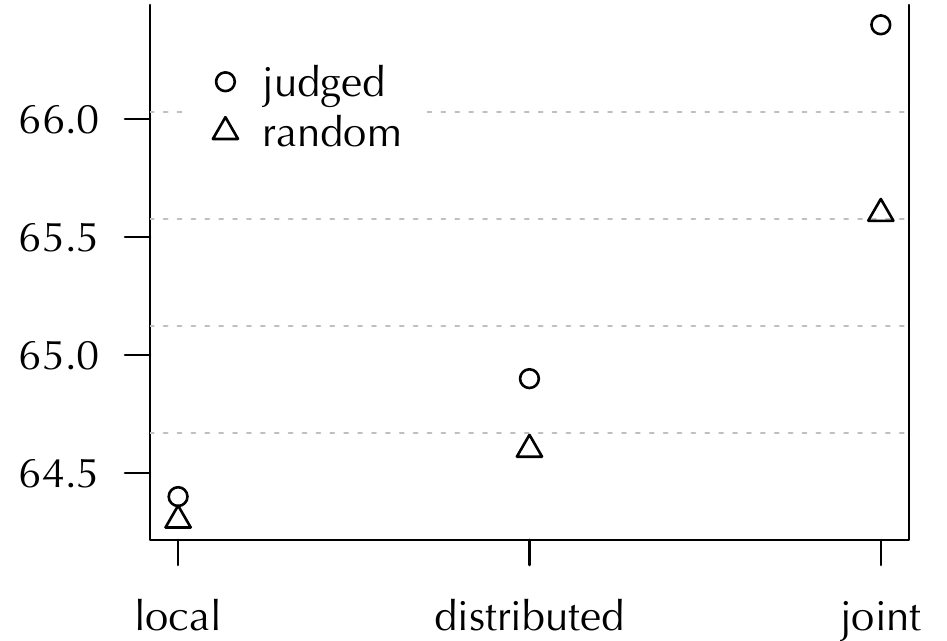}}
\caption{The duet model demonstrates significantly better performance ($p < 0.05$) on both test sets when trained with judged irrelevant documents as the negative examples, instead of randomly sampling them from the document corpus. The distributed model also shows statistically significant gain ($p < 0.05$) on the weighted set, and a non-statistically significant gain on the unweighted set.}
\label{fig:negsampling}
\end{figure}

We were interested in testing our hypotheses that training with labeled negative documents is superior to training with randomly sampled documents presumed to be negative.  We conducted an experiment training with negative documents following each of the two protocols. Figure \ref{fig:negsampling} shows the results of these experiments.  We found that, across all of our models,  using judged nonrelevant documents was more effective than randomly sampling documents from the corpus and considering them as negative examples.

\section{Discussion}
\label{sec:discussion}

Our results demonstrated that our joint optimization of local and distributed models provides substantial improvement over all baselines.  Although the independent models were competitive with existing baselines, the combination provided a significant boost.  

We also confirmed that using judged negative documents should be used when available.  We speculate that training with topically-similar (but nonrelevant) documents allows the model to better discriminate between the confusable documents provided by an earlier retrieval stage.  This sort of staged ranking, first proposed by \citet{cambazoglu:staged-retrieval}, is now a common web search engine architecture.

\begin{figure}
\centering
\subfigure[Model performance by query length]{\includegraphics[width=1.5in]{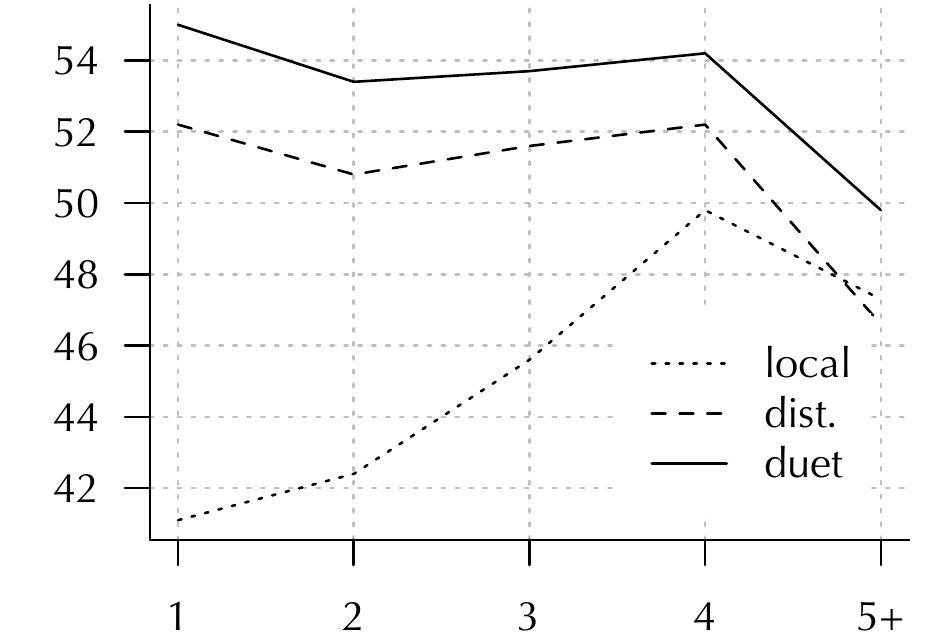}}\hfill\subfigure[Model performance by term rarity]{\includegraphics[width=1.5in]{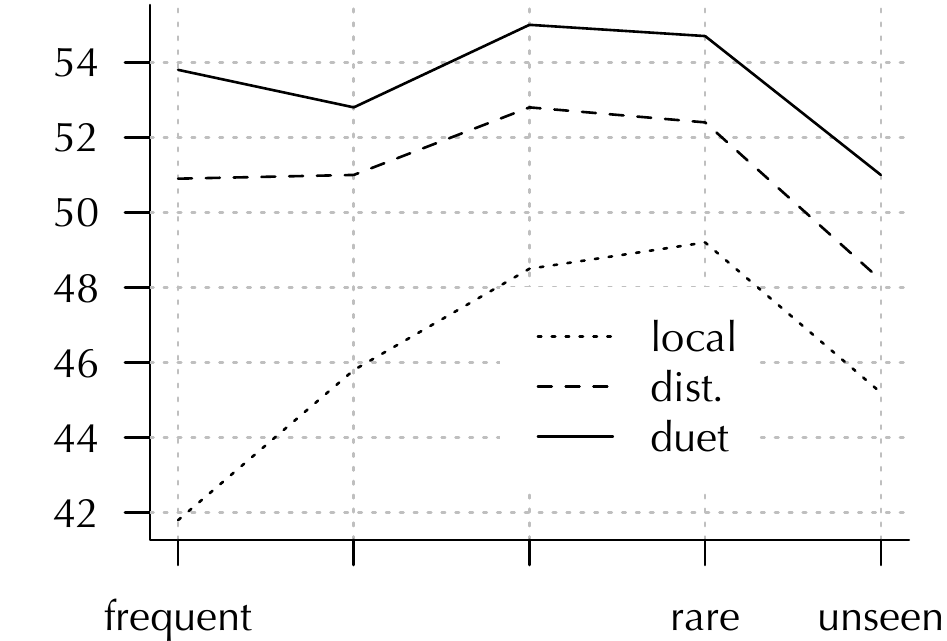}}
\caption{Performance of different models by length of query and how rare the rarest query term is in the training data. For the rare term analysis, we place all query terms into one of five categories based on their  occurrence counts in the training data. Then we then categorize each query in the test dataset based on the frequency of the rarest term belongs in the query. We include a category for queries with at least one term which has no occurrences in the training data.}
\label{fig:slicendice}
\end{figure}

In Section \ref{sec:baselines} we described our baseline models according to which of the properties of effective retrieval systems, that we outlined in Section \ref{sec:motivation}, they incorporate. It is reasonable to expect that models with certain properties are better suited to deal with certain segments of queries. For example, the relevant Web page for the query ``what channel are the seahawks on today'' may contain the name of the actual channel (e.g., ``ESPN'' or ``FOX'') and the actual date for the game, instead of the terms ``channel'' or ``today''. A retrieval model that only counts repetitions of query terms is likely to retrieve less relevant documents for this query -- compared to a model that considers ``ESPN'' and ``FOX'' to be relevant document terms. In contrast, the query ``pekarovic land company'', which may be considered as a tail navigational intent, is likely to be better served by a retrieval model that simply retrieves documents containing many matches for the term ``pekarovic''. A representation learning model is unlikely to have a good representation for this rare term, and therefore may be less equipped to retrieve the correct documents.
These anecdotal examples agree with the results in in Table \ref{tbl:results-main} that show that on the weighted test set all the neural models whose main focus is on learning distributed representations of text (duet model, distributed model, DESM, DSSM, and CDSSM) perform better than the models that only look at patterns of term matches (local model and DRMM). We believe that this is because the DNNs are able to learn better representations for more popular queries, and perform particularly well on this segment. Figure \ref{fig:slicendice} provides further evidence towards this hypothesis by demonstrating that the distributed model has a larger NDCG gap with the local model for queries containing more popular terms, and when the number of terms in the query is small. The duet model, however, is found to perform better than both the local and the distributed models across all these segments. 

\begin{figure}
\centering
\includegraphics[width=3in]{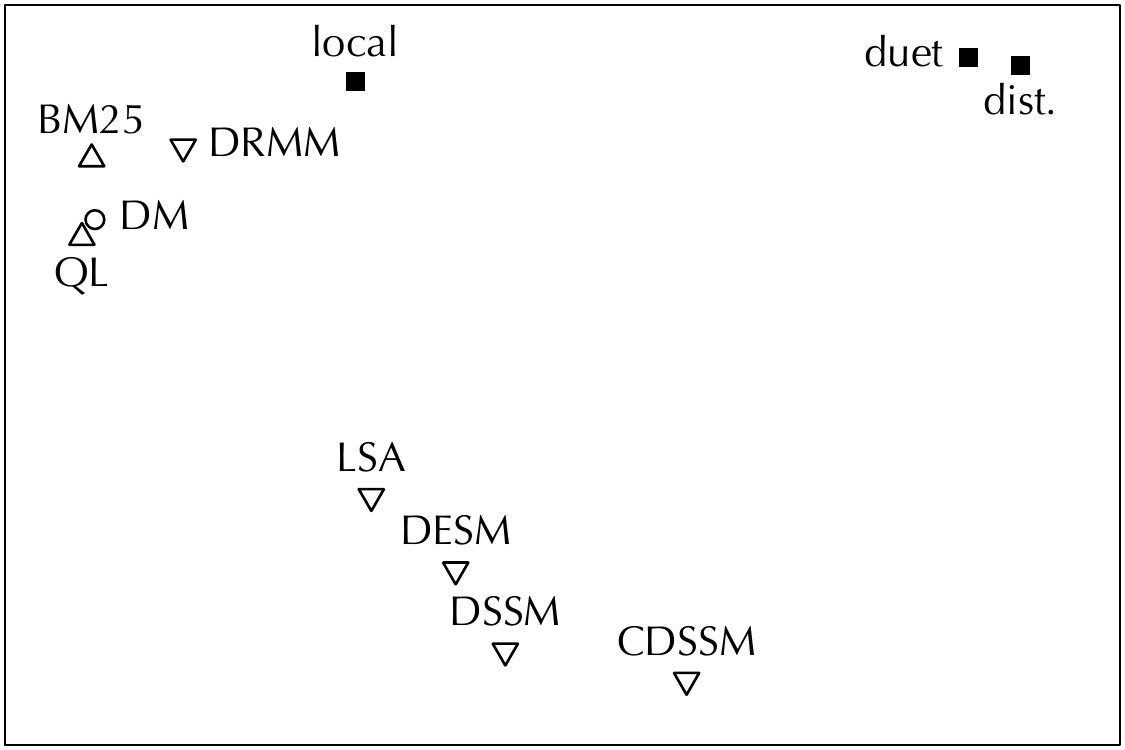}
\caption{Principal component analysis of models based on retrieval performance across testing queries.  Models using exact term matches ($\vartriangle$), proximity ($\circ$), and inexact matches ($\triangledown$) are presented.  Our models are presented as black squares.}
\label{fig:modelspace}
\end{figure}

In order to better understand the relationship of our models to existing baselines, we compared the per-query performance amongst all models.  We conjecture that similar models should performance similarly for the same queries.  We represented a retrieval model as a vector where each position of the vector contains the performance of the model on a different query. We randomly sample two thousand queries from our weighted test set and represent all ranking models as vectors of their NDCG values against these two thousand queries. We visualized the similarity between models by projecting using principal component analysis on the set of performance vectors.   The two-dimensional projection of this analysis is presented in Figure \ref{fig:modelspace}. The figure largely confirms our intuitions about properties of retrieval models. Models that use only local representation of terms are closer together in the projection, and further away from models that learn distributed representations of text. Interestingly, the plot does not distinguish between whether the underlying model is based on a neural network based or not, and only focuses on the retrieval properties of the model.

\begin{figure*}
\centering
\subfigure[Local model]{\includegraphics[scale=0.54]{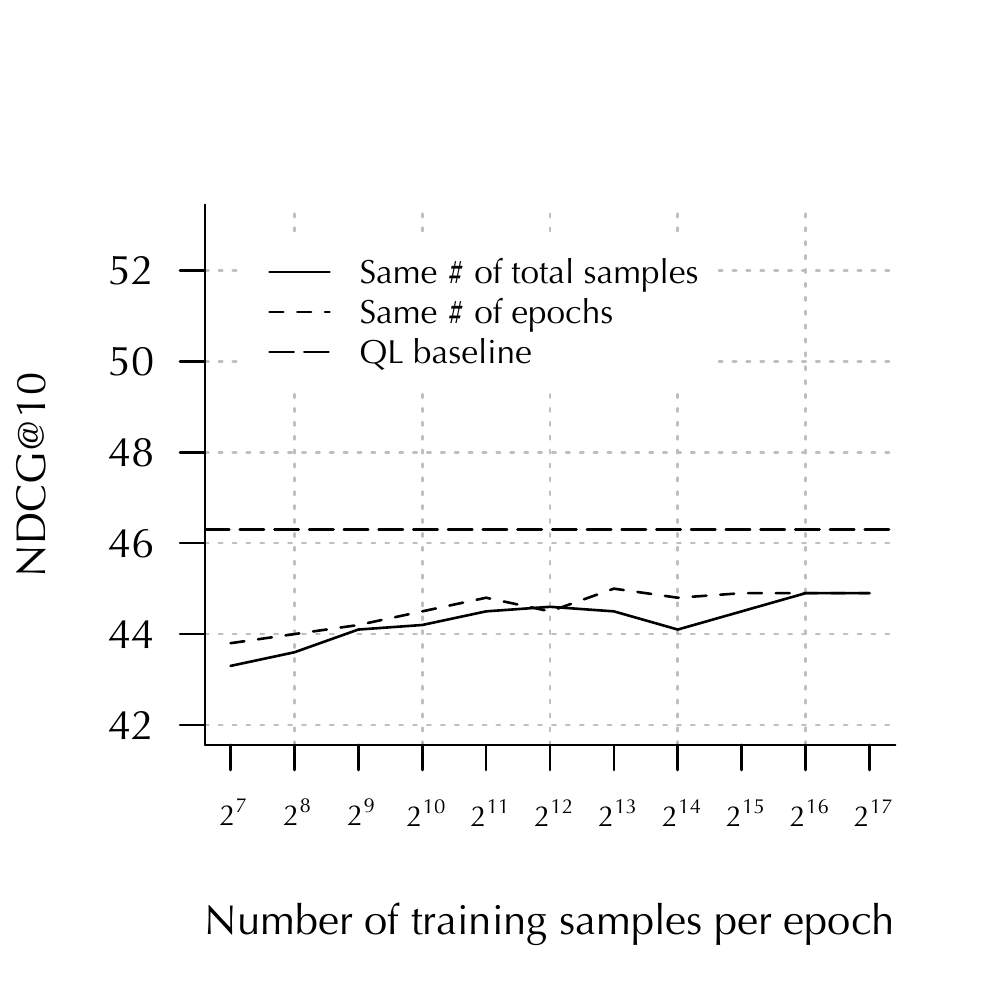}}
\subfigure[Distributed model]{\includegraphics[scale=0.54]{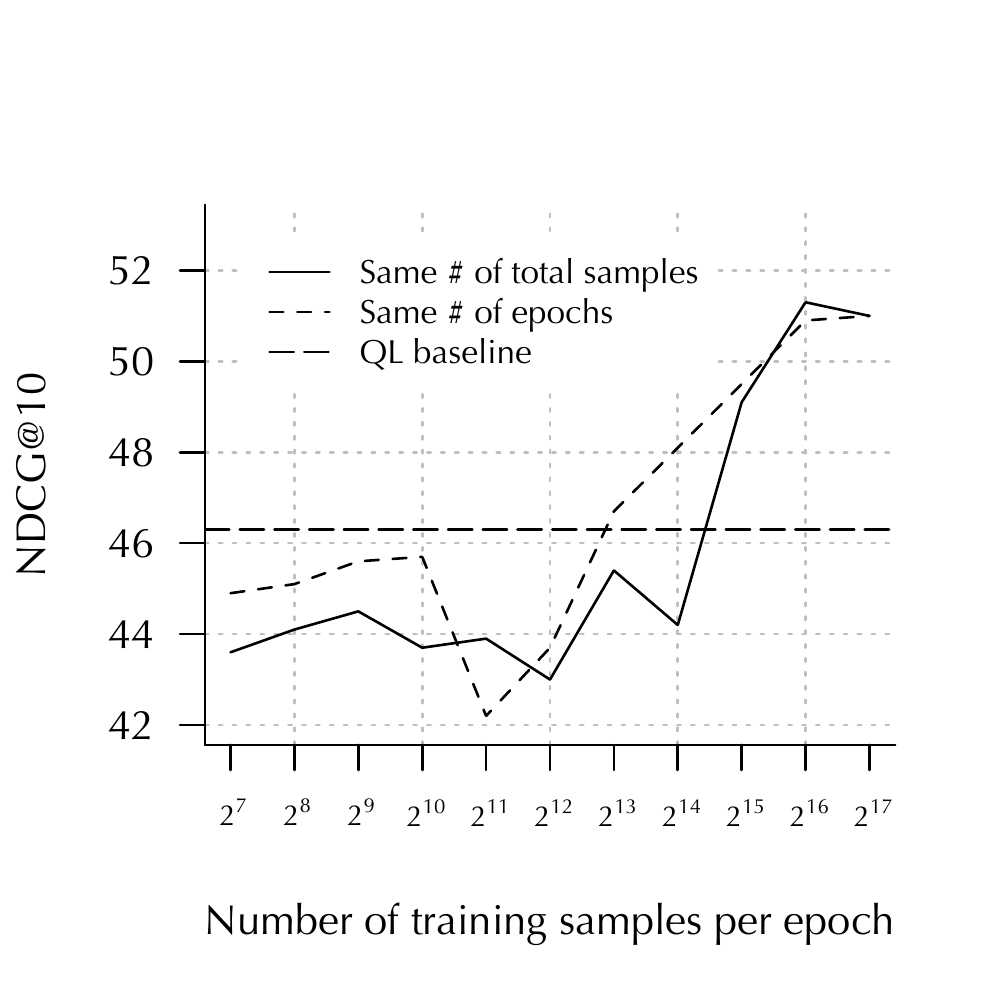}}
\subfigure[Duet model]{\includegraphics[scale=0.54]{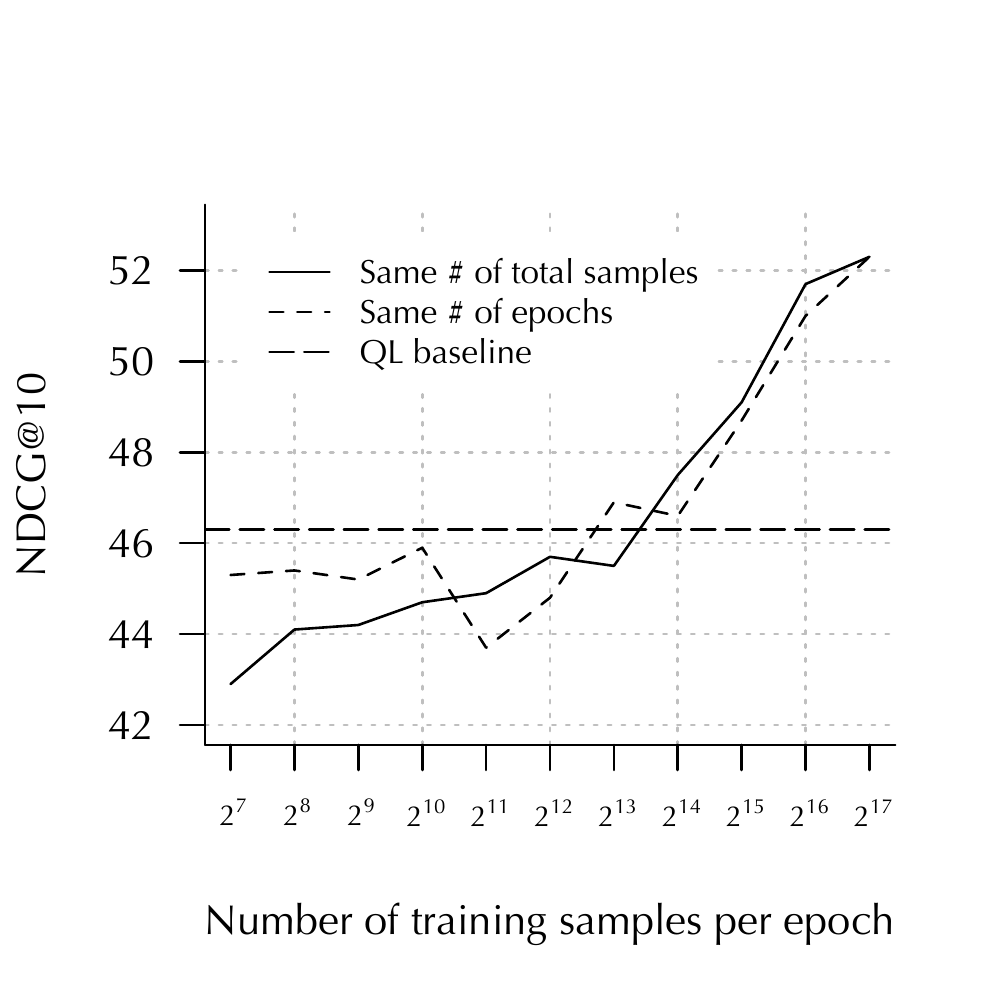}}
\caption{We study the performance of our model variants when trained with different size datasets. For every, dataset size we train two models -- one for exactly one epoch and another one with multiple epochs such that the total number of training samples seen by the model during training is 131,072.}
\label{fig:trainingdata}
\end{figure*}

Another interesting distinction between deep neural models and traditional approaches is the effect of the training data size on the performance of the model. BM25 has very few parameters and can be applied to new corpus or task with almost no training. On the other hand, DNNs like ours demonstrate significant improvements when trained with larger datasets. Figure \ref{fig:trainingdata} shows that the effect of training data size particularly pronounced for the duet and the distributed models that learns representations of text. The trends in these plots indicate that training on even larger datasets may result in further improvements in model performance over what is reported in this paper. We believe this should be a promising direction for future work.

A last consideration when comparing these models is runtime efficiency. Web search engines receive tens of thousands of queries per second. Running a deep neural model on raw body text at that scale is a hard problem. The local sub-network of our model operates on the term interaction matrix that should be reasonable to generate using an inverted index. For the distributed model, it is important to note that the $300 \times 899$ dimensional matrix representation of the document, that is used to compute the Hadamard product with the query, can be pre-computed and stored as part of the document cache. At runtime, only the Hadamard product and the subsequent part of the network needs to be executed. Such caching strategies, if employed effectively, can mitigate large part of the runtime cost of running a DNN based document ranking model at scale.

\section{Related Work}
\label{sec:related}

Representations of data can be local or distributed. In a local representation a single unit represents an entity, for example there is a particular memory cell that represents the concept of a grandmother. That cell should be active if and only if the concept of a grandmother is present. By contrast, in a distributed representation, the concept of grandmother would be represented by a pattern of active cells. \citet{hinton1984distributed} provides an overview contrasting distributed and local representations, listing their good and bad points. In a distributed representation, an activation pattern that has some errors or other differences from past data can still be mapped to the entity in question and to related entities, using a similarity function. A local representation lacks this robustness to noise and ability to generalize, but is better at precisely storing a large set of data.

This paper considers local and distributed representations of queries and documents for use in Web page ranking. Our measure of ranking quality is NDCG \cite{JK2002}, which rewards a ranker for returning documents with higher gain nearer to the top, where gain is determined according to labels from human relevance assessors. We describe different ranking methods in terms of their representations and how this should help them achieve good NDCG.

Exact term matching models such as BM25 \cite{robertson2009probabilistic} and query likelihood \cite{ponte1998language} tend to rank a document higher if it has a greater number of query term matches, while also potentially employing a variety of smoothing, weighting and normalization approaches. Such exact matching is done with a local representation of terms. Exact match systems do not depend on a large training set, since they do not need to learn a distributed representation of queries and documents.  They are useful in cases where the relevant documents contain exactly the query terms entered by the user, including very rare or new vocabulary, since new terms can be incorporated with no adjustments to the underlying model. They can also be extended to reward matches of query phrases and proximity \cite{metzler2005markov}.

To deal with the vocabulary mismatch problem that arises with local representations, it is possible to do document ranking using a distributed representation of terms. \citet{mikolov2013distributed} developed the popular word2vec embedding approach that has been used in a number of retrieval studies.
\citet{zheng2015learning} use term embeddings as evidence for term weighting, learning regression models to optimize weighting in a language modeling and a BM25 retrieval model.
\citet{ganguly2015word} used term embeddings for smoothing in the language modeling approach of information retrieval.
\citet{nalisnick2016improving} used dual embeddings, one for document terms and one for query terms, then ranked according to the all-pairs similarity between vectors. \citet{diaz2016query} used term embeddings to generate query expansion candidates in the language modeling retrieval framework, also finding better performance when training a specialized term embedding.

Other papers incorporating word embeddings include \cite{grbovic2015context, grbovic2015search, roy2016using}.

\citet{pang2016text} propose the use of matching matrices to represent the similarity of short texts, then apply a convolutional neural network inspired by those in computer vision. They populate the matching matrix using both local and distributed term representations. In the local representation, an exact match is used to generate binary indicators of whether the $i$th term of one text and $j$th term of the other are the same, as in our local model. In the distributed representation, a pre-trained term embedding is used instead, populating the match matrix with cosine or inner product similarities. The method works for some problems with short text, but not for document ranking \cite{pang2016study}. However, by using the match matrix to generate summary statistics it is possible to make the method work well \cite{guodeep}, which is our DRMM baseline.

These term embeddings are a learned representation of language, but in most cases they are not learned on query-document relevance labels. More often they are trained based on a a corpus, where a term's representation is learned from its surrounding terms or other document context. The alternative, learning a representation based on NDCG labels, is in keeping with recent progress in deep learning. Deep models have multiple layers that learn distributed representations with multiple levels of abstraction. This kind of representation learning, along with other factors such as the availability of large labeled data sets, has yielded performance improvements on a variety of tasks such as speech recognition, visual object recognition and object detection \cite{lecun2015deep}.

This paper learns a text representation end-to-end based on query-document ranking labels. This has not been done often in related work with document body text, but we can point to related papers that use short text such as title, for document ranking or related tasks. \citet{huang2013learning} learn a distributed representation of query and title, for document ranking. The input representation is character trigraphs, the training procedure asks the model to rank clicked titles over randomly chosen titles, and the test metric is NDCG with human labels. \citet{shen2014latent} developed a convolutional version of the model. These are our DSSM and CDSSM baselines. Other convolutional models that match short texts using distributed representations include \cite{hu2014convolutional,severyn2015learning}, also showing good performance on short text ranking tasks.

Outside of document ranking, learning text representations for the target task has been explored in the context of other IR scenarios, including query classification \cite{liu2015representation}, query auto-completion \cite{mitra2015suffixrank}, next query prediction \cite{sordoni2015hierarchical, mitraexploring}, and entity extraction \cite{gao2014modeling}.

\section{Conclusion}
\label{sec:conclusion}

We propose a novel document ranking model composed of two separate deep neural networks, one that matches using a local representation of text, and another that learns a distributed representation before matching. The duet of these two neural networks demonstrated a higher performance than the solo models on the document ranking task as well as  significant improvements over all baselines, including both traditional IR baselines and other recently proposed models based on shallow and deep neural networks. Our analysis indicate that these models may achieve even more substantial improvements in the future with much larger datasets.

\paragraph*{Acknowledgements}
The authors are grateful to Abdelrahman Mohamed, Pushmeet Kohli, Emine Yilmaz, Filip Radlinski, David Barber, David Hawking, and Milad Shokouhi for the insightful discussions and feedback during the course of this work, and to Frank Seide and Dong Yu for their incredible support with CNTK.

\setlength{\bibsep}{4pt}
\bibliographystyle{abbrvnat}
\raggedright{
\bibliography{bibtex}
}

\end{document}